\documentclass[]{aastex}
\usepackage[]{emulateapj5}
\usepackage[]{epsfig}
\usepackage{graphicx}

\hoffset -1.5truecm \lefthead{Giallongo et al.}
\righthead{}
\def\newline{\hfil\break}
\begin{document}
\title{AGNs as main contributors to the UV ionizing emissivity at high redshifts: predictions from a $\Lambda$-CDM model with linked AGN/galaxy evolution}
\author{E. Giallongo, N. Menci, F. Fiore, M. Castellano,  A. Fontana, A. Grazian, L. Pentericci}
\affil{INAF - Osservatorio Astronomico di Roma, via di Frascati
33, I-00040 Monteporzio, Italy}

\smallskip

\begin{abstract}
We have evaluated the contribution of the AGN population to the ionization history of the Universe based on a semi-analytic model of galaxy formation and evolution in the CDM cosmological scenario. The model connects the growth of black holes and of the ensuing AGN activity to galaxy interactions. In the model we have included a self consistent physical description of the escape of ionizing UV photons; this is based on the blast-wave model for the AGN feedback we developed in a previous paper to explain the distribution of hydrogen column densities in AGNs of various redshifts and luminosities, due to absorption by the host galaxy gas. The model predicts UV luminosity functions for AGNs which are in good agreement with those derived from the observations especially at low and intermediate redshifts ($z\sim 3$). At higher redshifts ($z>5$) the model tends to overestimate the data at faint luminosities. Critical biases both in the data and in the model are discussed to explain such apparent discrepancies. The predicted hydrogen photoionization rate as a function of redshift is found to be consistent with that derived from the observations. All that suggests to reconsider the role of the AGNs as the main driver of the ionization history of the Universe.

\end{abstract}

\keywords{galaxies: active --- galaxies: formation --- galaxies: evolution}

\section{Introduction}

The assessment of the thermal history of the intergalactic medium at high redshifts is fundamental to understand the physical processes involved in galaxy formation and evolution, including triggering and flueling of Active Galactic Nuclei (AGNs).

The ionization state of the intergalactic medium (IGM) is a function of the hydrogen ionizing UV background (UVB) conceivably produced by sources like high redshift galaxies and AGNs. Star forming galaxies are more common in the high redshift universe, and in principle can be responsible for the reionization of the IGM starting at $z>7$. However, even a first order estimate of their contribution to the ionizing UVB has two main sources of uncertainty. The first is connected with the increasing difficulty in evaluating the faint end slope of the UV luminosity function especially at the highest redshifts \citep{bouwens11,grazian11}. The second is even less understood and is connected with the poor knowledge of the average escape fraction of ionizing Lyman continuum photons from the interstellar medium of each galaxy. On the other hand, the AGN contribution is mainly affected by the poor knowledge of the faint end slope of their luminosity function especially at high redshifts due the the lack of deep AGN surveys at various wavelengths (see e.g. \citet{shankar07}).
The observed escape fraction of ionizing photons in the majority of the bright AGNs is high and of the order of unity at the Lyman continuum, although the value for the faintest AGNs is unknown.

Since the apparent number density of {\it bright} QSOs and AGNs is rapidly decreasing outside the redshift interval $2<z<3$ it is usually assumed that star forming galaxies can give a considerable contribution to the ionizing flux which could become dominant at $z<1$ and at $z>3$ \citep{madau91,giallo97,hama96,hama12}.
In this context several attempts have been made to derive the escape fraction of UV ionizing photons both at low redshifts from space and at high redshifts $z=2-4$ from ground based observations. However, there is little evidence supporting a scenario where enough ionizing photons escape from galaxies at intermediate redshifts, and most attempts gave only upper limits on the escape fractions.

Indeed recent measurements by \cite{cowie09} provided a stringent $2\sigma$ upper limit of 0.8\% for the relative escape fraction of galaxies at $0.9<z<1.4$ suggesting that galaxies are not the dominant contributors to the ionizing UV background at $z<2$. At high redshifts the situation is more difficult from the observational point of view. Direct measures of the Lyman continuum flux from galaxies at $z>4$ are increasingly difficult because of the sharp increase in the number density of intervening Lyman Limit absorptions due to optically thick clouds in the IGM. For this reason the search has been focused to the range $z\sim 3-4$, but there a discrepancy is found by different teams using different methods. Steidel and collaborators support the detection of appreciable Lyman limit flux escaping at least from the UV brightest galaxies at $z\sim 3$ \citep{steidel01}. More recent analyses suggest average escape fractions of order $10-15$\%
\citep{shapley06,nestor11}, although their sample can be contaminated by foreground objects \citep{vanzella10,boutsia11}.
Other teams using both spectroscopic and very deep broad or narrow band imaging gave only upper limits in the range $<5-15$\% \citep{giallo02, fernandez03,vanzella10b,boutsia11}.
To summarize, the recent trends appear to favor low escape fractions from relatively bright galaxies, where the recent limits are below 1\% at low redshifts and below 10\% at $z\sim 3-4$, and cast serious doubts on any redshift evolution in the galaxy ionizing escape fraction. This leaves open the possibility that AGNs could still play a leading role in contributing to the cosmological ionizing background. 

The traditional view based on relatively bright QSO surveys  in optical and X-ray bands lead to the standard view of an evolutionary ionizing QSO background increasing from the local value out to $z=2-3$ and then quickly decreasing (see e.g. \citet{madau99}). Previous estimates were based on simple parametric extrapolations both in luminosity and redshifts of the AGN luminosity function without any link to specific physical processes \citep{madau99}. However recent deep otpical QSO surveys at $z=3$ \citep{siana08,fontanot07} and $z=4$ \citep{glikman11} are showing the presence of a considerable number of previously unknown faint AGNs producing a rather steep luminosity function. The inclusion of X-ray detection in the selection methods enables to extend the knowledge of the luminosity function to even fainter limits \citep{fiore12} of the order of L(2-10 KeV) $> 10^{43}$ erg s$^{-1}$ out to $z> 4$.

The presence of this faint population is changing our estimate of the AGN contribution to the ionizing UV background although the selection of faint AGNs at the highest redshifts $z\sim 6$ becomes difficult with the current instrumentation. Given the current observational limits, an insight into the origin 
of the ionizing flux at high redshifts can be gained from theoretical modelling. The latter should account for the formation and evolution of galaxies in a cosmological context, and for the co-evolution of the AGN population. Given the complex sub-grid physics involved in the AGN activity over a wide range of cosmic times and of galactic masses, semi-analytic models (SAMs) constitute a powerful tool  to provide a statistically relevant sample of simulated 
galaxies and AGNs. Such models connect the physical processes involving gas and star formation to the merging histories of DM haloes
collapsed from the primordial density field. While several SAMs including the growth of supermassive Black Holes have been presented in the literature (starting from \citet{kauffmann00}; see also \citet{croton06,bower06,marulli08,menci06,menci08}), addressing the problem of the ionizing background requires a physical description of the escape fraction in galaxies. 
Such a problem is deeply connected with that concerning the feedback of the AGNs onto the interstellar gas. In fact, since even a tiny amount (column densities $N_H\sim 10^{17}$ cm$^{-2}$) of galactic gas 
would suppress the radiation at the Lyman limit escaping from the galaxy, estimates of the escape fraction translate into computing the fraction of 
photons that are emitted along directions where the interstellar gas of the galaxy has been depleted. In our previous work \citep{menci08}, we have developed our SAM model
(Rome-SAM, hereafter R-SAM)) to describe the expansion of the blast wave produced in the galaxy interstellar medium by AGN-driven outflows. Here we use the same model to derive the escape fraction of photons in active galaxies as the fraction of photons emitted along 
directions where the AGN-induced blast wave has expelled the galactic gas. When coupled with the integrated luminosity function of AGN self-consistently derived in the same model, this allows us to compute the ionizing background produced by AGNs within a cosmological model of galaxy formation. 
We stress that the basic  quantities predicted by our computation have been extensively tested against observations in our previous works.
On the one hand, the predictions concerning the growth of supermassive black holes and the corresponding evolution of the luminosity function have been 
compared against  a wide range of different observations, from optical to X-rays, for  SMBHs masses and AGN luminosities in a wide range of cosmic times extending up to z=6 for the brightest QSOs (\citet{menci03,menci04,menci06,menci08}; see also \citet{fiore12}); on the other hand, the blast-wave model for the AGN feedback allowed us to derive -within a full cosmological context and in connection with the galaxy properties- the distributions of column densities in active galaxies, as a function of redshift and of the AGN luminosity,  which we successfully tested against observations \citep{menci08}. 
In this paper we extend the analysis of the UV AGN luminosity function to faint objects and adopt the blast-wave mechanism to predict the amount of UV photons escaping from the AGN host galaxies.

The paper is organized as follows: in Sect. 2.1 we recall the basic properties of our semi-analytic model for the evolution of galaxies and AGNs. 
In Sect. 2.2 we describe how our blast-wave model for the AGN feedback (developed in previous papers) allows to compute the luminosity-dependent 
escape fraction of ionizing photons emitted by AGN. The evolution of the luminosity function of AGNs predicted by our model is compared with 
data in Sect. 3; the comparison with the current dataset is  discussed taking into consideration possible biases in the model and in the data.
In Sect. 4 we convolve the predicted intrinsic luminosity function of the AGN with the escape fraction derived in our model to derive the emissivity and the 
photoionization rate predicted by the model at various redshifts, and compare the latter with existing data; the contribution of AGNs with different 
luminosities to such observables is also shown and discussed. The final Sect. 5 is devoted to a summary and to conclusions. 

\section{AGNs emission in Hierarchical Galaxy Evolution}

The semi-analytic model R-SAM we develop and use  connects the AGN evolution with the  evolution of galaxies in a cosmological framework (see \citet{menci08} and references therein).

\subsection{Hierarchical Galaxy/AGN Evolution in the R-SAM}

Galaxy formation and evolution is driven by the collapse and growth of dark
matter (DM) haloes, which originate by gravitational instability of  overdense
regions in the primordial DM density field. This is taken to be a random,
Gaussian  density field with Cold Dark Matter (CDM) power spectrum expected within the
''concordance cosmology" \citep{spergel07} for which we adopt round
parameters  $\Omega_{\Lambda}=0.7$, $\Omega_{0}=0.3$, baryonic density
$\Omega_b=0.04$ and Hubble constant $h=0.7$ in units of 100 km/s/Mpc. The
normalization of the spectrum is taken to be $\sigma_8=0.9$ in terms of the variance
of the field smoothed over regions of 8 $h^{-1}$ Mpc.
The merging histories of DM haloes and sub-haloes is followed through a Monte Carlo simulation of the collapse and subsequent merging history of the peaks of the primordial density field, which enables us to generate a synthetic catalogue of model galaxies and of their past merging history.
The physics of baryons (gas radiative cooling, disk formation, quiescent and impulsive star formation, stellar and AGN feedback) is connected to the merging histories of DM haloes as described in detail in our previous papers (e.g., \citet{menci05}). Note that
the impulsive star formation implemented in the  model is triggered not only by (major and minor) merging events but
also by fly-by events between galaxies.
The R-SAM also includes a treatment of SMBHs growing at the centre of galaxies
by interaction-triggered inflow of cold gas, following the physical model proposed
by \citet{cavaliere00} and implemented in \citet{menci03,menci06,menci08}.
The accretion of cold gas is triggered by galaxy encounters (both of fly-by and
of merging kind), which destabilize part of the galactic cold gas mass $m_c$  by inducing
loss af angular momentum.
The fraction of cold gas accreted by the BH in an interaction event is computed
in terms of the variation $\Delta j$ of the specific angular momentum $j\approx
Gm/v_d$ of the gas, to read \citep{menci03} $f_{acc}\approx 10^{-1}\,
\big|{\Delta j/ j}\big|$.
For minor merging events and for the encounters among galaxies with very unequal mass ratios
$m'/m\ll 1$, that dominate the statistics in all hierarchical models of galaxy formation, the
accreted fraction takes on values $10^{-3}\lesssim f_{acc}\lesssim 10^{-2}$.
The average amount of cold gas accreted during an accretion episode is thus
$\Delta m_{acc}=f_{acc}\,m_c$, and the duration of an accretion episode, which provides
the timescale for the QSO or AGN to shine, is assumed to be the  crossing time
$\tau$ of the destabilized cold gas component.
The  time-averaged bolometric luminosity so produced by a QSO hosted in a given galaxy
is then provided  by
\begin{equation}
L={\eta\,c^2\Delta m_{acc}\over \tau} ~.
\end{equation}
We adopt an energy-conversion efficiency $\eta= 0.1$ (see \citet{yu02}), and derive luminosities in the various bands on adopting
standard spectral energy distributions as in \citet{marconi04} and Marconi et al. (in preparation). The SMBH mass $m_{BH}$ grows mainly through accretion
episodes as described above, besides  coalescence with other SMBHs
during galaxy merging. As initial condition, we assume small seed
BHs of mass $10^2\,M_{\odot}$ \citep{madau01} to be initially
present in all galaxy progenitors; our results are insensitive to
the specific value as long as it is smaller than some
$10^5\,M_{\odot}$.
Due to the crucial role of AGN feedback in affecting the star formation history of massive galaxies, this part of the model has been extensive tested. The predicted QSO luminosity function and X-ray luminosity function and emissivity have been checked against observation over a wide range of redshift $0 < z < 5$ and bolometric luminosities $10^{43} < L  < 10^{46}$ erg s$^{-1}$ \citep{menci03,menci04}. The consistency of the modeling of the AGN feedback with the colour distribution of galaxies, including a description of the bimodal appearance, has been discussed in \citet{menci06}.

\subsection{Blast-Wave Model for AGN feedback and the escape fraction of UV ionizing radiation}

Fast winds with velocities up to $10^{-1}c$ are observed in the central regions of AGNs. They likely originate from the acceleration of disk outflows by the AGN radiation field (see \citet{begelman03} for a review), and affect the environment in the host galaxy and beyond, leaving imprints out to large scales of some $10^2$ kpc in the intracluster medium (ICM).
Clear examples of outflows in local AGNs are those observed on galactic scales in the infrared and optical bands in Mrk 231 and Mrk 573 \citep{feruglio10,fischer10,rupke11}.
A detailed model for the transport of energy from the inner, outflow region to larger scales has been developed by \citet{cavaliere02} and \citet{lapi05}. Central, highly supersonic outflows compress the gas into a blast wave terminated by a leading shock front, which moves outwards with a lower but still supersonic speed and sweeps out the surrounding medium. 
The key quantity determining all shock properties is the total energy $\Delta E$ injected by AGNs into the surrounding gas. This is computed as

\begin{equation}
\Delta E=\epsilon _{AGN} \eta c^2 \Delta m_{acc}=\epsilon _{AGN}\,L\,\tau
\end{equation}

for each SMBH accretion episode in our Monte Carlo simulation; the value of the energy feedback efficiency for coupling with the surrounding gas is taken as $\epsilon _{AGN}=5\times 10^{-2}$, which is consistent with the values required to match the X-ray properties of the ICM in clusters of galaxies (see \citet{cavaliere02}). This is also consistent with the observations of wind speeds up to $v_w\sim 0.1c$ in the central regions, which yield
$\epsilon _{AGN} \sim v_w/2c\sim 0.05$ by momentum conservation between photons and particles (see \citet{chartas02}; \citet{pounds03}); this value has also been adopted in a number of simulations (e.g., \citet{dimatteo05}) and in our R-SAM (e.g.,\citet{menci06}).

The blast expands into the ISM or the ICM as described by hydrodynamical equations that include the effects not only of an initial density gradient, but also those of upstream pressure and DM gravity, clearly important quantities in galaxies, as discussed by \citet{lapi05}. The solutions show in detail how the perturbed gas is confined to an expanding shell bounded by an outer shock at radius $R_s(t)$, which sweeps out the gas surrounding the AGN. 
An analytic expression for $R_s(t)$ has been derived by the above authors in the case of shock expansion in a gas with power-law density profile
$\rho\propto r^{-w}$; in the relevant central regions of the galaxies (within $1/2$ the scale-length $r_d$ of the exponential disk) 
the expansion of the shock is given by (see \citet{menci08}) 
\begin{equation}
R_s(t)\approx r_d\,{\mathcal M} \,t/t_d
\end{equation}
where ${\mathcal M}$ is the Mach number. This is related to the energy $\Delta E$ (eq. 2) injected by the AGN (relative to the initial thermal energy content 
$E\propto m_c$ of the galactic gas) by the relation  ${\mathcal M}^2=1+\Delta E / E$ \citep{lapi05}. Thus the expansion velocity of the bast wave is directly related  to the AGN luminosity.

The above blast-wave model for the AGN feedback allows to self-consistently compute the escape fraction of ionizing photons. In fact, 
the typical galactic column densities are in the range $N_H\approx 10^{20}\div 10^{24}$ cm$^{-2}$ and effectively suppress the escape of ionizing 
radiation along the line-of-sights where the galactic gas has not been swept out by the blast wave produced by the AGN. Thus, the escape fraction 
can be computed as $f(L) =\omega /2\,\pi$, where $\omega$ is the solid angle subtending  the intercept of the expanding blast wave with the 
galactic disk. This is illustrated in fig. 1, which shows how only ionizing photons propagating along the directions (indicated with 
$\ell_1$ in the figure) within the solid angle $\omega$ can travel unabsorbed and escape the galactic disk, while directions outside the 
solid angle $\omega$ (indicated  with $\ell_2$ in the figure) intercept a fraction of the gas disk thus blocking the escape of ionizing photons. 
Note that the solid angle $\omega$ increases as the shock radius $R(t)$ propagates outward.

We can compute the solid angle $\omega=2\pi\,(1-cos\,\theta/2)$ in terms of the opening angle $\theta$ (the aperture of the inner cones in fig. 1), which in turn is 
related to the vertical scale height $h$ of the disk and to the radius of the expanding shell $R(t)$ by the relation $cos\, \theta/2=h/R(t)$. 
Thus we obtain the time-dependent escape fraction $f(L,t)=\omega/2\pi=1-h/R(t)$, where the shock radius $R(t)$ depends on the AGN luminosity 
(see eq. 3 and text below).  Since we are interested in the statistical, effective escape fraction, we compute the average over the duration $\tau$ of the 
AGN activity
\begin{equation}
\langle f(L)\rangle={1\over \tau}\,\int_{t_S}^{\tau}\,dt\,\Bigg[1-{h\over R(t)}\Bigg].
\end{equation}
Here $t_S$ is the time (computed from the start of the AGN phase) at which the shock radius first encompasses the width of the disk (for earlier times 
escape fraction is null),  so that $R(t_S)=h$. Using eq. (3) we express $t_S$ as a function of the disk dynamical time $t_d$ as  $\tilde{t}_S\equiv t_S/t_d=(h/r_d)/\mathcal M $. Performing the integral in eq. (4) then yields
\begin{equation}
\langle f(L)\rangle=1- { \tilde{t}_S \over \tilde{\tau} }\Bigg[1+ { \ln{\tilde{t}_S \over \tilde{\tau}}}\Bigg], 
\end{equation}
where $ \tilde{\tau}=\tau/t_d\approx 1$ (for interaction driven AGNs). Note that for shocks with $\mathcal M\gg 1$ (i.e., large AGN injected energies
$\Delta E\gg E$, see eqs. 2 and 3)  imply $\tilde{t}_S\ll \tilde{\tau}$, yielding 
large average escape fractions $\langle f(L)\rangle\approx 1$, while slowly expanding shocks yield  $\langle f(L)\rangle\approx 0$. Thus eq. (5) results in an  
effective escape fraction increasing with the AGN luminosity $L$. At fixed $L$ the escape fractions in eq. (5) increase with redshift since - at given $\Delta E$  - the smaller host galaxy masses yield larger $\Delta E/E$ at higher $z$.

For a  ratio $h/r_d=1/15$ (see \citet{narayan02}) determining the timescale $t_S$ for each AGN luminosity (corresponding to $h\approx 200$ pc for a typical $L_∗$ galaxy), we obtain low-redshift ($z\simeq 0.5$) escape fractions increasing from $\langle f\rangle\approx 0.1$ for $M_{1450}=-22$ to $\langle f\rangle\approx 0.4$ at 
$M_{1450}=-24$, while at higher redshifts ($z\simeq 4$) the values range from $\langle f\rangle\approx 0.3$ for $M_{1450}=-22$ to $\langle f\rangle\approx 0.8$ at 
$M_{1450}=-24$; quantitative informations on the dependence of the escape fraction on the AGN luminosity and redshift can be found in fig. 2. 
Note that the value of the parameter $h/r_d=1/15$ adopted here to compute $\langle f\rangle$ from eq. (5)  
is  the same adopted in \citet{menci08}, where 
we showed that the blast-wave model for AGN feedback provides column densities $N_H$ (computed after averaging over all the line of sights 
schematically shown in fig. 1) depending inversely on the AGN luminosity. Such a behaviour is due to the fact  faster  shocks correspond to a lower fraction of still unperturbed gas outwards of $r=R_s (t)$, and is in quantitative agreement with observations, as we showed in the paper mentioned above.  

\section{The AGN UV luminosity functions}

To predict the UV ionizing emissivity produced by our AGN evolution model we first evaluate the evolution of the AGN luminosity function as a function of redshift at $\lambda=1450$ {\AA}, where most of the UV rest frame data on AGNs are collected and where the average AGN emission shows the peak in the spectral energy distribution, representative of the overall bolometric emission. The contribution of each luminosity to the ionizing background 
will be the computed by convolving such luminosity functions with the escape fraction in eq. (5).

In Fig. 2 we show the evolution of  the UV luminosity function $N(L_{1450})$ computed in our R-SAM (binned in magnitude, dashed line), and the fraction $N(L_{1450})\,\langle f\rangle$  that will contribute to the ionizing background (solid line), where $ \langle f\rangle$  is the luminosity-dependent  escape fraction given in eq. (5). 
The bolometric luminosity $L$ (computed for each AGN in our R-SAM after eq. 2) is related to the  emission $L_{1450}$ at 1450 $\AA$ after the bolometric correction given in \citet{marconi04} and Marconi et al. (in preparation).

In the figure we compare our predictions with data from literature. Note that the data should be compared with the global UV luminosity function (dashed line). At  redshifts $z\leq 2.5$ we have compared our predictions with the luminosity function provided by \citet{richards09} and \citet{croom09}. The data come from the combined 2dF-SDSS sample of Quasars photometrically selected from Sloan Digital Sky Survey, and then spectroscopically confirmed using the 2DF instrument on the Anglo Australian Telescope. The luminosity functions are complete down to $M_{1450}=-21.5$ and show a gradual flattening below $M_{1450}=-24$. The shaded region brackets the determinations derived in the X-ray band 
by \citet{lafranca05,ebrero09,aird10}, and thus
 accounts for systematic errors only. Conversions to 1450 {\AA} have been obtained using the \citet{marconi04} relation. Note that \citet{lafranca05,ebrero09} determinations account for the obscured sources whose flux is reduced below the flux limit by
intervening column densities, they provide densities slighly higher
than those of Aird et al. 2010, which are not corrected for this
effect.

In this redshift range, 
the agreement with the model predictions is good,  if we consider that the errors shown are only Poissonian and that various systematics  (e.g. in the flux corrections) probably would widen the uncertainties involved. In any case, the luminosity functions derived from X-ray surveys (see shaded regions in fig 2) constitute upper boundaries that are fully consistent with our  predictions. 

At higher redshifts $z> 3$ we compare with the rest-frame UV luminosity functions from \citet{fontanot07}, \citet{bongiorno07}, \citet{siana08} and \citet{glikman11}. Here the comparison with data at the faint end is more critical, since the corrections for incompleteness of the observations play a more relevant role.
For example, the data by \citet{siana08}  have been derived combining IR SPITZER SWIRE  data and optical through a standard morphological criterion and a color selection based on the UV dropout expected at these redshifts coupled with an IR color selection aimed at removing low redshift interlopers. 
These authors attain a completeness level at the faintest survey limits of about 75\% which is taken into account in their effective volumes. However, their completeness estimates are based on detailed simulations involving both templates of quasar spectra and a model for IGM absorption, which is crucial for any UV dropout selection. In their analysis they refer to the \citet{madau95} model, whose analysis has been superseded since larger statistical information about the IGM absorption became available (e.g. \citet{meiksin06}, \citet{inoue08}), so that now the Madau model represents an upper limit to the average absorption. More recent and reliable estimates by \citet{prochaska09} show average Lyman limit opacities by the IGM observed in 1800 Sloan quasars at $z\sim 3.5-4.5$  about a factor two lower. This should allow the presence of a significant number of $z\sim 3$ quasars bluer than expected by the UV dropout threshold adopted by \citet{siana08}, implying a lower completeness level of the luminosity function difficult to quantify. A similar bias could be present in the \citet{fontanot07} data which go two magnitudes fainter at a slightly higher average redshift, $z\sim 3.7$. The luminosity function has been derived from combined standard color selection and X-ray matching in the GOODS fields. The color selection was effective in the interval $3.5<z<5.2$. In Fig.2 we show the luminosity function derived in the $3.5<z<4$ interval where the completeness of the survey is higher. Even in this survey the average IGM absorption adopted to derive the color threshold is similar to that predicted by the Madau model at least up to $z\sim 4.5$, although they apply some correction factor to the Lyman $\alpha$ forest. The important point in this respect is that at $z>3$ the incidence of IGM absorption systems which dominate the Lyman limit opacity has a slower evolution compared to the Lyman $\alpha$ lines \citep{prochaska09}. A further important source of incompleteness at  magnitudes $M_{1450}\gtrsim-24$ is constituted by the morphological selection criterion adopted to reduce in the spectroscopic follow up the contamination from  Lyman break galaxies at the same redshift as described in \citet{glikman10,glikman11}.

In the same fig. 2 we include two faint points derived from a recent very deep search for AGNs in the GOODS south field by \citet{fiore12}. The latter authors adopt different criteria starting from the standard color dropout technique but including faint detection in the  4 Ms Chandra images in the 2-10 KeV X-ray band and relaxing in this way any morphological restriction. They use spectroscopic redshifts where available or accurate photometric redshifts from the GOODS MUSIC and GOODS ERS multicolor catalogs \citep{grazian06,grazian11}. In this way they can select faint AGN activity among the numerous Lyman break galaxies in a wide redshift interval $3<z<7$ putting interesting constraints on the faint end of the luminosity function at $z=3-4, 4-5$, and $z>5.8$ dominated by Seyfert-like objects.  Although a fraction of the order of $20$\% of the detected AGNs appears strongly absorbed in X-ray or Compton thick, the same objects do not show any reddening in the optical-UV part of the spectrum which shows emission lines with various widths. Indeed subsequent spectroscopic optical follow up revealed even in the case of Compton thick sources, typical highly ionized UV emission lines like CIV and NV. In summary, the limited spectroscopic information available coupled with the analysis of the overall spectral energy distributions are suggesting the absence of any significant link between the presence of strong absorption in X-ray and emission of UV photons from the central nucleus, at least in cases where the X-ray flux is measured for optically-UV selected AGNs.
In this respect the sample is more complete respect to the optical surveys and this is reflected in the higher densities attained by the \citet{fiore12} luminosity functions. Nevertheless the sample could be affected by several uncertainties and significant incompleteness, especially at the faintest limits where the objects have been selected from very deep HST NIR images at $H\sim 27$ and measured at the faintest X-ray fluxes of $F_X\sim 2\times 10^{-17}$ ergs s$^{-1}$ cm$^{-2}$. At the lowest X-ray luminosities, objects with relatively high X/Optical ratio, and consequently with $H>27$, are missed in their survey. Most of the redshifts have been evaluated photometrically and this could introduce low redshift interlopers in the sample. On the other hand, even in this case the color selection based on the standard UV dropout technique adopts a predicted IGM opacity higher than measured by \citet{prochaska09} although lower than expected in the Madau model, allowing for some incompleteness. As a consequence uncertainties in the Fiore et al. LF data due to systematic errors could be larger than shown in the figure based on number statistics. Keeping in mind these limitations we have adapted their LF data after having recast the X-ray luminosities in UV (1450 {\AA}) magnitudes following the standard \citet{marconi04} corrections.

At redshifts $z>4$ the above observational biases and possible incompleteness could explain 
the model over prediction appearing at fainter magnitudes $M_{1450}>-24$ respect to the present dataset. The discrepancy is a factor $2-3$ in the magnitude interval $-24<M_{1450}<-22$ up to $z\sim 6$ and increases to fainter magnitudes. Since the estimate of the observational incompleteness is difficult to assess without the data input from deep  X-ray surveys, we shall rely on our model to get an insight on the contribution of AGNs to the ionizing background. To this aim, we shall couple our predicted luminosity functions with 
the physical description of the escape fraction based on the blast-wave model. 

\section{The AGNs Lyman Continuum Emissivity and Hydrogen Photoionization Rate}

We now proceed to compute the contribution of our predicted luminosity functions to the AGN emissivity at the Lyman limit. This is given 
by 
\begin{equation}
 \epsilon =\int \langle f(L)\rangle N(L)\,L\,\,{E_{912}\over E}\,dL 
\end{equation}
where the  $\langle f(L)\rangle$ is the fraction of AGNs of a given luminosity which shows Lyman Continuum emission given by eq. (5), 
and $E/E_{912}$ is the (luminosity dependent) bolometric correction at 912 {\AA}. In the integral we have considered all the AGNs produced in our MonteCarlo code down to very faint magnitudes, however the total emissivity at a given redshift is independent of the faintest luminosity cutoff. This is 
clearly shown in fig. 3, which shows the contribution to the total emissivity from AGNs with different bolometric luminosity $L$ at different redshifts. 
Since the bolometric correction from the UV $M_{1450}$ magnitude is not linear, we have indicated on top of the first panel the UV magnitudes at 1450 {\AA} corresponding to the peak bolometric luminosities. 
Note that while at $z\sim 3$ the main contribution comes from AGNs in the relatively wide magnitude interval $-24<M_{1450}<-18.5$, at higher redshifts the main contribution comes from brighter AGNs with $-24<M_{1450}<-22$. In this magnitude interval the discrepancy between data and model prediction is within a factor $2-3$, which is of the same order as the differences among surveys where AGNs are selected with different methods in various bands.

In this respect it is just in this magnitude interval that a more careful analysis
should be performed to investigate whether data and model prediction can be reconciled. Unfortunately this magnitude interval is sparsely sampled by present surveys; more statistics with much more controlled systematics should be produced in the future.
In particular, multiwavelength surveys of medium depth (e.g. $I\sim 22$) in an adequate sky area should be planned for spectroscopic and X-ray follow up. In the meanwhile we explore the ability of our linked AGN/galaxy model in providing sufficient UV photons to keep the IGM ionized up to $z\sim 6$. \textbf{In fig.4 and Table 1 the predicted ionizing UV emissivity  $ \epsilon_{24} $ in units of $10^{24}$ erg s$^{-1}$ Hz$^{-1}$ Mpc$^{-3}$ is shown as a function of redshift. The sharp increase up to $z\sim 2$ is due to the luminosity evolution of the AGN luminosity function. At $z>4.5$ the emissivity gradually declines up to $z\sim 8$ and then drops quickly. The rather flat behaviour at high redshifts is due to the frequent galaxy merging rate which drives the AGN activity coupled with the short gas cooling time. Both guarantee a high AGN duty cycle and consequently an average AGN emissivity weakly evolving in time.
}

The photoionization rate per hydrogen atom $\Gamma_{-12}$ in units of $10^{-12}$ s$^{-1}$ is the standard quantity \textbf{used to evaluate the ionization status of the IGM}. We have computed this quantity following \citet{madau99} and \citet{schirber03}
\begin{equation}
\Gamma_{-12}(z)\simeq 0.46 {\epsilon_{24}(z)\over 3+\mid \alpha_{UV}\mid} \left(\Delta l\over 50Mpc\right){\left(1+z\over 4.5\right)}^{3-(2.5+\gamma)}
\end{equation}
where $ \alpha_{UV}=-1.76 $ is the average UV spectral index of AGNs derived from the average SED we adopt \citep{marconi04}. The power-law index $-(2.5+\gamma)$ describes the decrease in redshift of the mean free path (mfp) $\Delta l$ of ionizing photons in the IGM due to the increase in redshift of the Lyman Limit absorption systems with slope $\gamma$. We adopt $\gamma=1.94$ and the normalization of the mfp to 50 Mpc at $z=3.5$ as in \citet{songaila10}.  At $z> 4.2$ we adopt a steeper redshift evolution with $\gamma=5$ to fit the mfp data point at $z\simeq 5.8$ (see \citet{songaila10} for details). This equation represents a first order approximation where only ionizing sources within one absorption length (where $ \tau_{IGM} =1 $) contribute to the ionizing background.
The resulting curve from our MonteCarlo model is shown in Fig.5. The dotted one represents the intrinsic production of UV ionizing photons by the overall AGN population. The ratio between the two curves depends on the ratio between the Lyman limit optically thin AGNs and the overall population.

Data points with different symbols are also shown for comparison. They represent different estimates derived in a model-dependent way from the IGM statistics.
Indeed most values come from methods of estimating the UV background; and so $\Gamma$ 
based on the mean flux decrement in the Lyman $\alpha$ forest of QSO spectra in combination with numerical simulations. The UV background value is adjusted until the mean flux in simulated Ly$\alpha$ forest spectra is equal to that in real data
\citep{faucher08,bolton05,becker07,fan06,wyithe11}.
An alternative method is based on the proximity effect which has recently been proposed at the highest redshifts $z=5-6$ by \citet{calverley11}. However
both methods are subject to uncertainties and systematics which are
responsible for the scatter among the data points reported in Fig. 5. The data shown in the figure derived from flux decrement analysis in the IGM Lyman forest of QSO spectra, have been scaled to the same IGM temperature-density relation as in \citet{calverley11}.

Indeed, converting the mean flux decrement into an ionization rate depends on modelling the gas density and temperature distribution in the IGM at very low densities \citep{miralda00} in various cosmological scenarios. On the other hand, estimates from the proximity effect are sensistive to the assumption concerning the presence of overdensities of matter close to the measured quasar and to the assumption that gas temperature within the proximity region is similar to that in the general IGM \citep{calverley11}.
 
The continuous curve represents the hydrogen ionization rate as a function of redshift 
predicted by our model, where the oscillating behaviour is related to the stochastic nature of the AGN trigger, and rendered by our MonteCarlo code. The model is able to reach an almost constant value of the order of $\Gamma_{-12} \simeq 1$ consistent with the data points, in a broad redshift interval from $z\sim 2$ up to $z\sim 4$. A decline by a factor 4 is then appearing at higher redshifts up to $z\sim 6$ mainly due to the significant decrease of the mean free path of ionizing photons.


The gradual decline at high redshifts is a strong intrinsic prediction of our model based on a linked AGN/Galaxy evolution and  is in line with recent data analyses at $z\sim 5-6$ \citep{calverley11,wyithe11}. As we have shown in the previous section, the model overestimates the number of relatively ($M_{1450}\sim -23$) faint AGNs at $z>4$ by a factor $\sim 2-3$; on the other hand we argued the possible presence of significant incompleteness in the scanty present faint dataset.

Since the predicted emissivity by hard spectrum sources like AGNs keeps high up to $z\sim 6$ it is interesting to evaluate the impact of this evolution on the ionization history of HeII in the IGM. Indeed the higher HeII ionization threshold respect to hydrogen and its smaller photoionization cross-section imply a complete HeII reionization only when numerous hard spectrum sources like AGNs are distributed within the IGM.  There is some consensus about a late HeII reionization which could complete at $z\sim 3$ as suggested by observations of HeII Ly$\alpha$ Gunn-Peterson trough and patchy absorption in the HeII Ly$\alpha$ forest at $z\sim 3$ (\citet{shull10} and reference therein). Other indirect evidence comes e.g. from an apparent increase of the average IGM temperature at $z\sim 3$ measured from metal line ratios (e.g. \citet{songaila98}).

Following early analysis by \citet{madau99} and more recently by \citet{hama12} we have computed the volume filling factor of HeIII in the IGM as a function of redshift.  We have also considered the case where Lyman limit absorption by optically thick clouds (LLSs) limits the AGN ionization,
as proposed by \citet{bolton09}. In the latter case each AGN is able to ionize only a limited region
due to the fact that, at variance with hydrogen ionization, the distance between HeII Lyman limit optically thick clouds is smaller than the typical AGN separation at $z\gtrsim 3$. Thus  HeII-ionizing photons will have intersected several helium LLSs being subject to absorption. The equation describing the evolution of the HeIII volume filling factor $Q$ is, according to \citet{bolton09}: $\dot{Q} =\dot{n}-RQ-\dot{n}Q$ where $\dot{n}$ is the HeII ionization rate per helium atom and $R(t)$ is the average recombination rate which includes an effective clumping factor. The last term describes absorption by LLSs clumps. In Fig.6 we show the evolution of the volume filling factor assuming only recombination with moderate clumping factors. We also show the evolution including  LLSs absorption as a further photon sink as in \citet{bolton09}. The $Q$ evolution adopting the \citet{hama12} emissivity is also shown for comparison. It appears that the HeII reionization can be completed down to $z\sim 3$ assuming modest clumping especially when LLSs allow each AGN to ionize only a limited volume. The early hydrogen reionization is shown for comparison only in the recombination case since the mfp of hydrogen LLSs is larger than typical AGN separation and the photon sink term can be neglected. Of course the knowledge of the detailed HeII reionization history requires 3D hydrodynamical models. The adopted simplified description is only suggesting that some degree of inhomogeneity could help to reduce the tension between the possible presence of a high level of ionizing emissivity by AGNs at $z\sim 6$ with an HeII reionization ending at $z\sim 3$.

\section{Summary and Conclusions}

A faint AGN population is emerging from recent multiwavelength surveys, in particular from those including X-ray detections. These are changing our perception about the AGN contribution to the ionizing UV background, although the selection of faint AGNs at the highest redshifts $z\sim 5-6$ becomes difficult with the present instrumentation. Given the current observational limits,  insight into the origin of ionizing flux at high redshifts can be gained from theoretical modelling. We have adopted our semianalytic model R-SAM which successfully links the evolution of the galaxy population to that of the AGN population through the computation of the growth of supermassive black holes in the nuclei of galaxies. The model was able to reproduce several observables such as luminosity and mass functions of both galaxies and AGNs.

In this context we first concentrated on the comparison between the predicted and observed LFs at the UV rest-frame wavelength $1450$ {\AA}. The agreement is satisfactory at low and intermediate redshifts up to $z\sim 3$. At higher redshifts the predicted LFs tend to overestimate the observed data by a factor $2-3$ at the fainter magnitudes $M_{1450}\sim -22$. We have discussed the possible reasons for the discrepancy. From the observational point of view it is clear that surveys of optically/UV selected AGNs which also rely on X-ray detection as a discriminant in the candidate selection process appear more complete on approaching the predicted values by the model. Indeed without the X-ray discriminant,
the point-like morphological criterion is needed to avoid confusion with the more numerous Lyman break galaxy population. But this introduces significant incompleteness at faint magnitudes. Moreover several uncertainties in the selection process like e.g. the lack of spectroscopic information for the faintest AGN candidates, make the estimates of the faint end of the AGN LF appreciably uncertain at very high redshifts.

Since faint AGN surveys could be affected by significant bias we have relied on our model to get an estimate of the hydrogen photoionization rate produced by the AGNs. To this aim, we have coupled our predicted luminosity functions with the physical description of the escape fraction based on the blast-wave model already introduced in our R-SAM to describe the expansion of the blast-wave produced in the interstellar medium of the hosting galaxy by the AGN-driven outflows. The blast wave clears the way for the UV photons emitted by the nuclear AGN which are free to escape outside the galaxy and ionize the intergalactic medium.
To evaluate whether AGNs can ionize the intergalactic medium we have first computed the AGN emissivity at the Lyman limit as a function of luminosity and redshift. Interestingly, the predicted AGN emissivity has a definite maximum at about $M_{1450}\sim -23$ which is almost independent of redshift. Fainter sources give a negligible contribution because of the flattening of the shape of the luminosity function and because the escape fraction of UV photons decreases on average in fainter AGNs and reaches $\langle f\rangle\approx 0.1-0.3$ at $M_{1450}\sim -22$ at intermediate  and high redshifts, respectively. Note that for AGNs of a given luminosity the escape fraction increases with redshift following the increase of the merging rate which is the main triggering of the QSO activity in our model. Since at $z>3$  AGNs with $M_{1450}\sim -22$ show a duty cycle of about 10\% and so represent  about 10\% of the host galaxy population \citep{fiore12} then the bulk of the Lyman Continuum emitters hiding faint AGN activity represents about 1\% of the UV-dropout galaxy population similar to recent statistics on large Lyman break galaxy samples \citep{vanzella10}.
The resulting photoionization rate evolves strongly from $z=0$ to $z=2$, and then remains nearly constant to a value  $\Gamma_{-12} \sim 1$ up to $z\sim 4.5$. A gradual decline is apparent at higher redshifts, consistent with the trend derived from most recent analyses of the Lyman forest in QSO spectra. We have also shown that  a high level of AGN ionizing emissivity up to $z\sim 6$ could still be consistent with a more gradual HeII reionization ending at $z\sim 3$, especially if absorption by HeII Lyman limit IGM clouds is taken into account.

In summary, the main prediction of the model is that in a scenario where the AGN activity is linked to the dynamical galaxy evolution triggered by interactions and merging, only a mild high redshift evolution of the hydrogen photoionization rate is expected in the redshift interval $z\sim 2-5$. New deep multiwavelength surveys with accurate selection procedures are needed to build reliable
AGN volume densities down to $M_{1450}\sim -23$ in the relevant redshift interval. These values are critical to understand whether AGNs can play a key role in reionizing the high redshift IGM.
Our model suggests that it is time to reconsider the AGNs as the main population driving the ionization history of the Universe over the whole known redshift interval (up to $z\sim 6$) where the cosmic evolution of structures is observed.

\acknowledgments

We thank the referee for useful comments which have significantly improved our paper. We are grateful to A. Cavaliere for critical reading of the manuscript. We also thank F. Haardt for useful discussion on the Helium reionization history.

\clearpage
Table 1.\\

\begin{tabular}{|l|r|l|r|}
\hline
redshift & $\log {\epsilon_{24}}$ & redshift & $\log {\epsilon_{24}}$\\
\hline
 0.25    & -0.66  &    4.75   &   1.16     \\
  0.50    & -0.23  &  5.00   &   1.14 \\
  0.75    &  0.06  &  5.25   &   1.12 \\
   1.00  &   0.30  & 5.50   &   1.10 \\
   1.25 &    0.46  &  5.75   &   1.08\\
   1.50  &   0.59  &   6.00   &   1.05\\
   1.75   &  0.66  &  6.25   &   1.03\\
   2.00    & 0.78  &  6.50   &   1.00 \\
   2.25   &  0.86  &  6.75  &   0.97\\
   2.50   &  0.91  &  7.00  &   0.94\\
   2.75   &  0.96  &  7.25  &   0.91\\
   3.00  &    1.01  &  7.50  &   0.87\\
   3.25 &     1.06  &  7.75  &   0.81\\
   3.50  &    1.10   & 8.00  &   0.75\\ 
   3.75   &   1.14   &  8.25  &   0.69\\
   4.00    &  1.17  &   8.50  &   0.58   \\
   4.25    &  1.19   &  8.75  &   0.46\\
   4.50   &   1.18  &    9.00   &  0.32\\
\hline
\end{tabular}

 \clearpage

\begin{center}
\vspace{0.cm}
\scalebox{0.57}[0.57]{\rotatebox{0}{\includegraphics{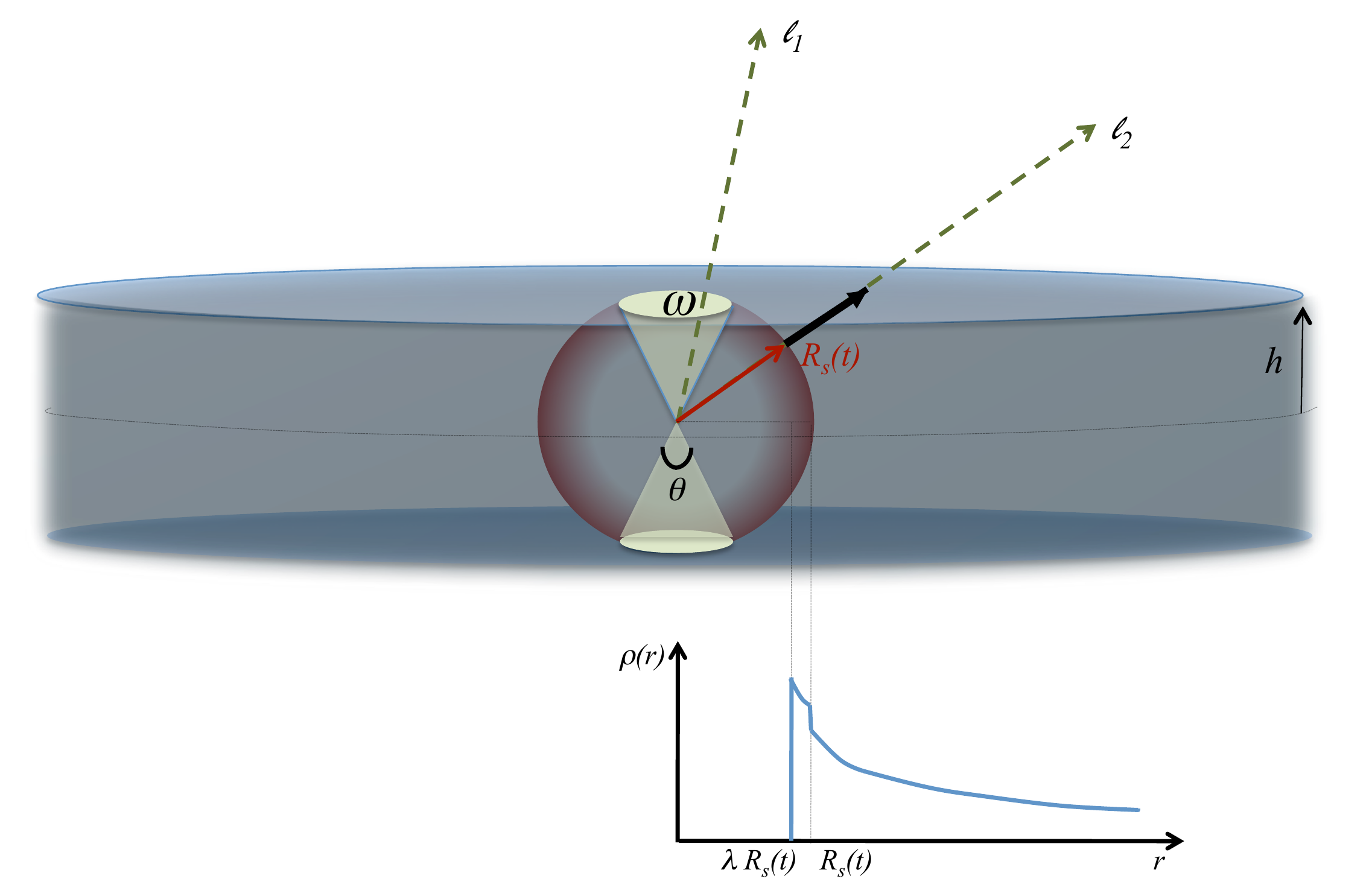}}}
\end{center} {\footnotesize \vspace{-0.4cm }
Fig. 1. - A schematic representation of the geometrical effects determining the 
escape fraction in the blast wave model of AGN feedback. The shock radius $R_s(t)$
expands outwards, compressing the swept gas into a thin shell
(represented in darker colour) with width $R_s(1-\lambda)$ (see \citet{lapi05} for a computation of $\lambda$), and
leaving a cavity inside. Such a density distribution $\rho(r)$ is
plotted in detail at the bottom. 
The solid angles $\omega$ subtend the intercept of the expanding blast wave with the 
galactic disk, and are represented as grey (yellow in the electronic paper) cones. 
Ionizing photons can travel unabsorbed only along the directions $\ell_1$ contained within $\omega$, 
while the directions external to the cones (indicated as $\ell_2$) intercept a fraction of the 
galactic gas, thus inhibiting the escape of ionizing photons.
 \vspace{0.2cm}}

\clearpage

\begin{center}
\vspace{-0.7cm}
\scalebox{0.6}[0.6]{\rotatebox{0}{\includegraphics{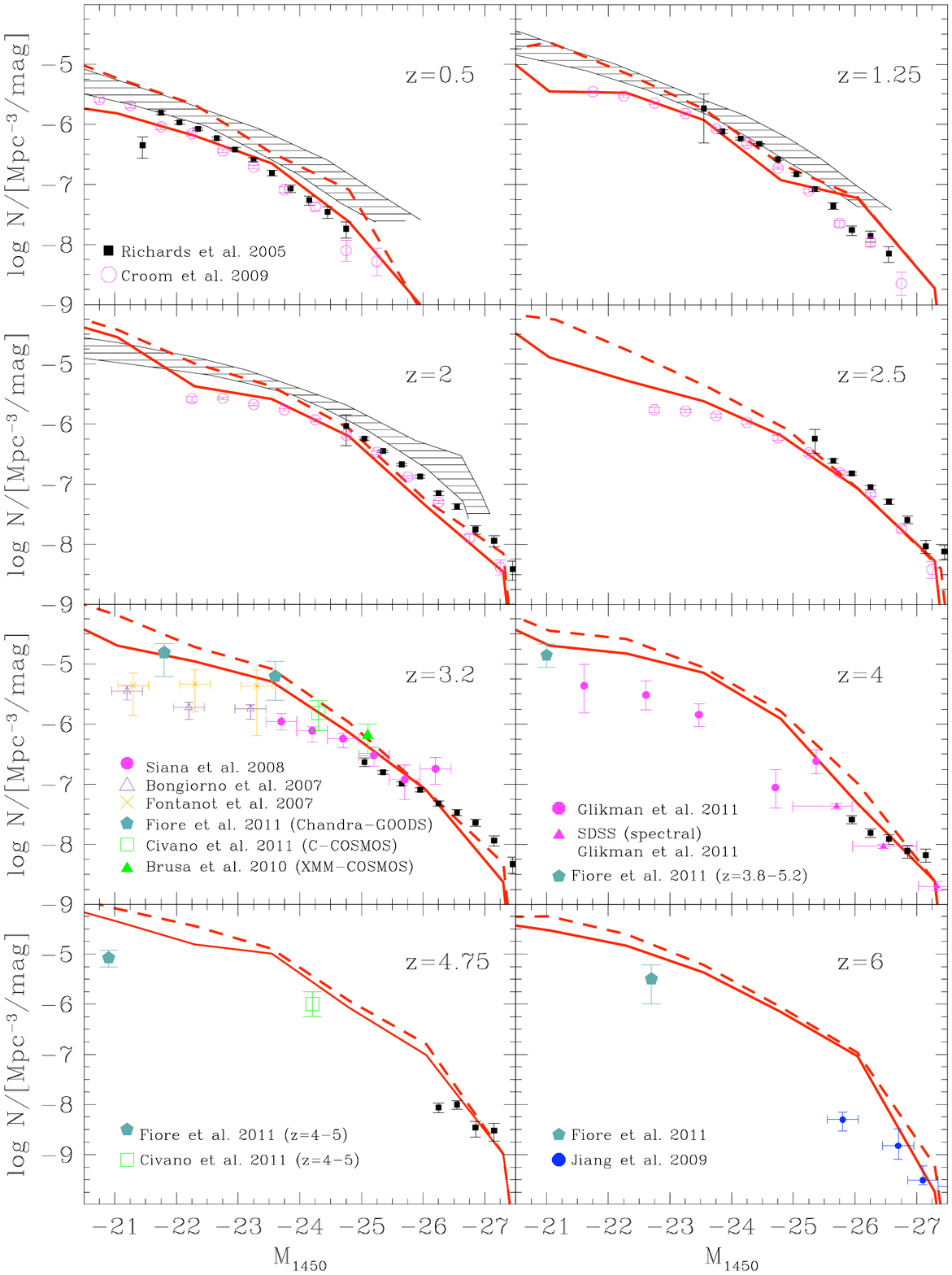}}}
\end{center} {\footnotesize \vspace{-1.5cm }
Fig. 2. - The evolution of the UV luminosity function  obtained from our SAM model (lines) is compared with different observational 
data. Dashed lines refer to the full intrinsic emission at 1450 {\AA}, while solid lines show the fraction of objects that will contribute to the 
ionizing background and suppress the luminosity function by the luminosity-dependent  escape fraction given in eq. (5). 
The redshift intervals and the data we compare with are shown in the different panels and discussed in the text. 
\vspace{0.2cm}}
 
\clearpage

\begin{center}
\vspace{-0.2cm}
\scalebox{0.52}[0.52]{\rotatebox{-90}{\includegraphics{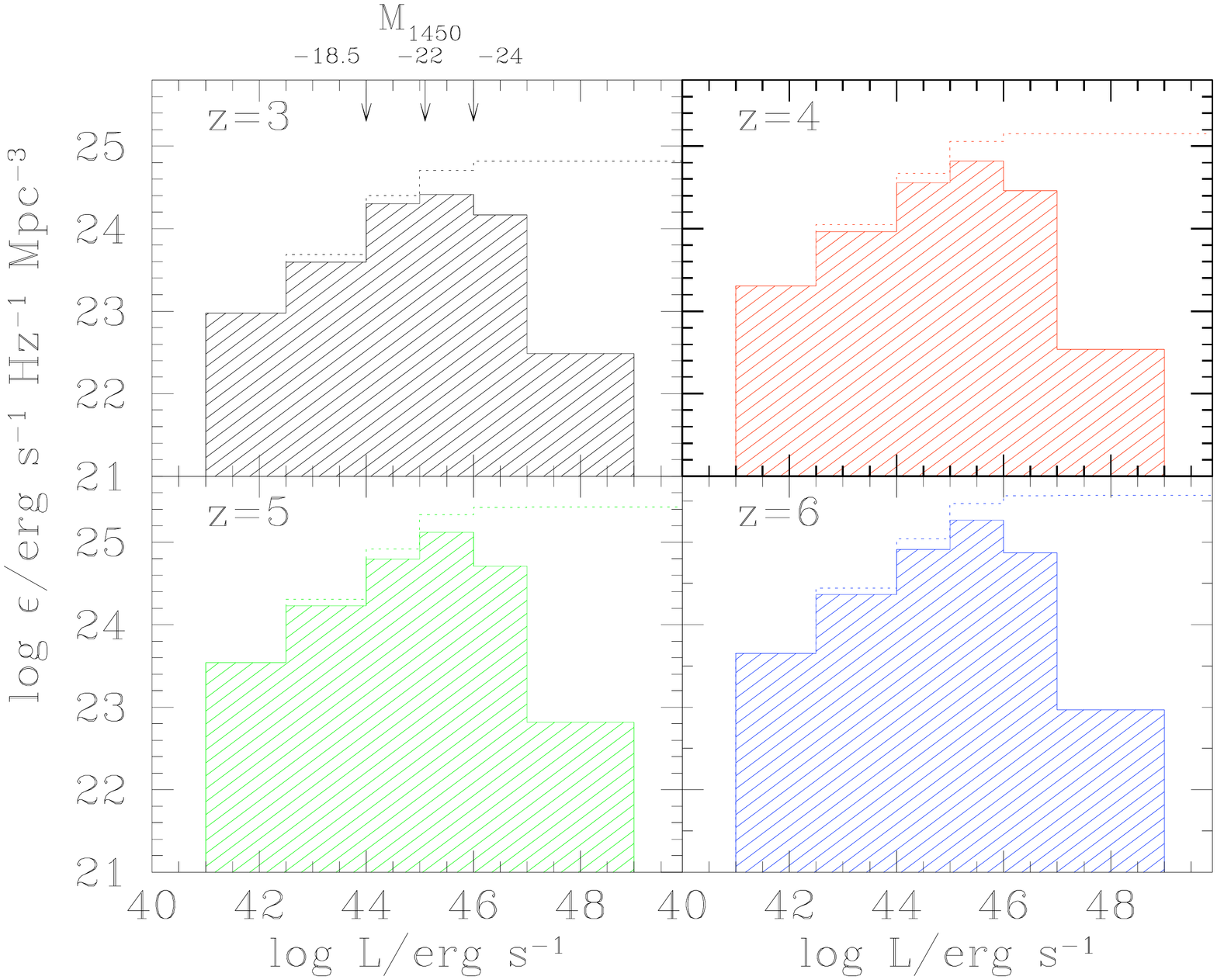}}}
\end{center} {\footnotesize \vspace{0.2cm }
Fig. 3. - The integrated emissivity of the AGN population (computed from eq. 6) contributed by AGNs with different bolometric luminosity (on the x-axis) 
in different redshift bins. The histograms represent the differential contribution, while the cumulative contribution is represented by the dotted line. 
The  UV magnitudes at 1450 {\AA} corresponding to the peak bolometric luminosities are also shown on the scale at the top of the top-left panel. 
 \vspace{0.4cm}}

 \clearpage
 
 \begin{center}
\vspace{-0.5cm}
\scalebox{0.5}[0.5]{\rotatebox{0}{\includegraphics{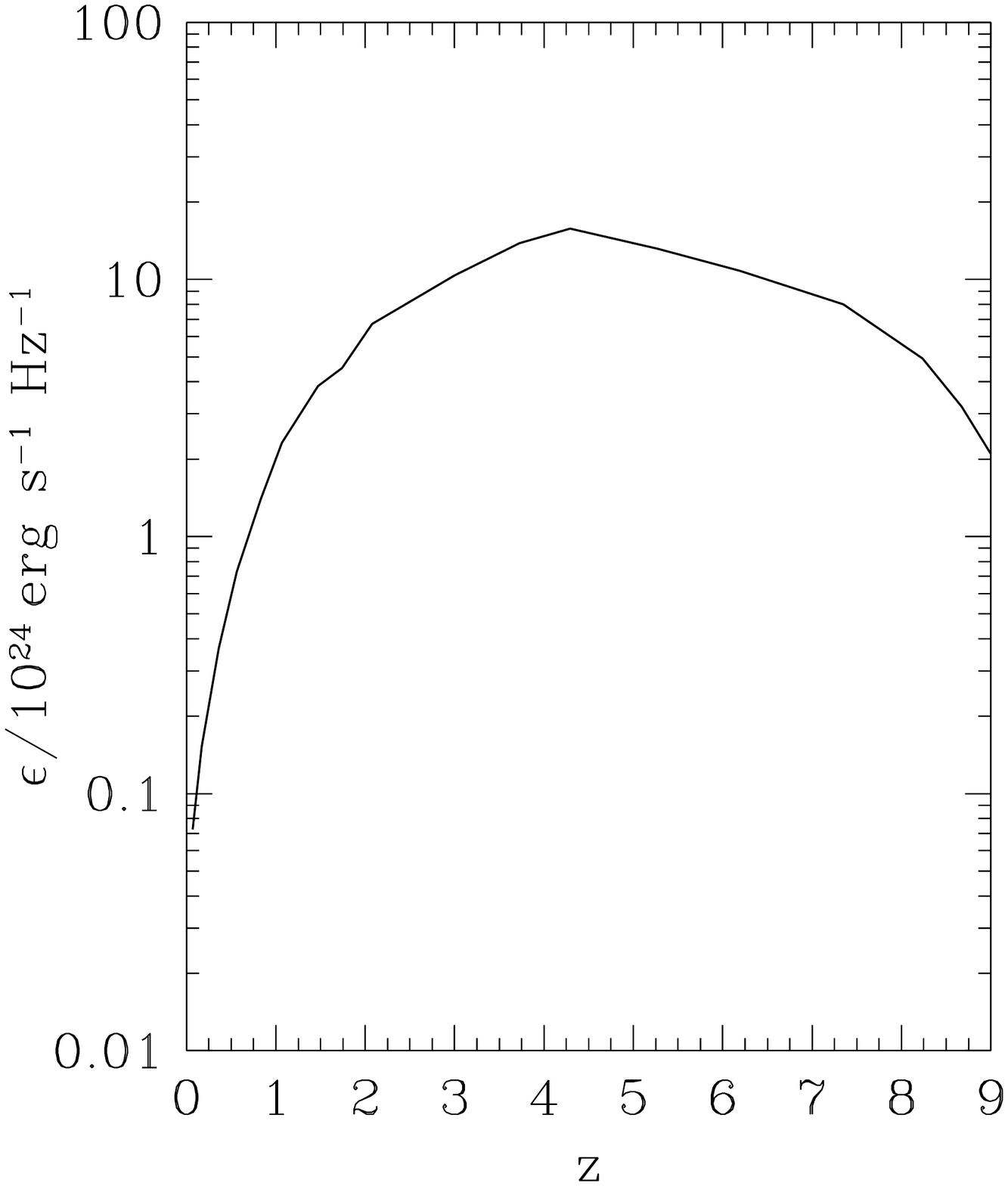}}}
\end{center} {\footnotesize \vspace{-0.2cm }
Fig. 4. - Lyman continuum emissivity produced by the AGN population in the R-SAM model as a function of redshift. The curve has been smoothed respect to the original MonteCarlo values.
 
\clearpage 
 
\begin{center}
\vspace{-0.5cm}
\scalebox{0.5}[0.5]{\rotatebox{0}{\includegraphics{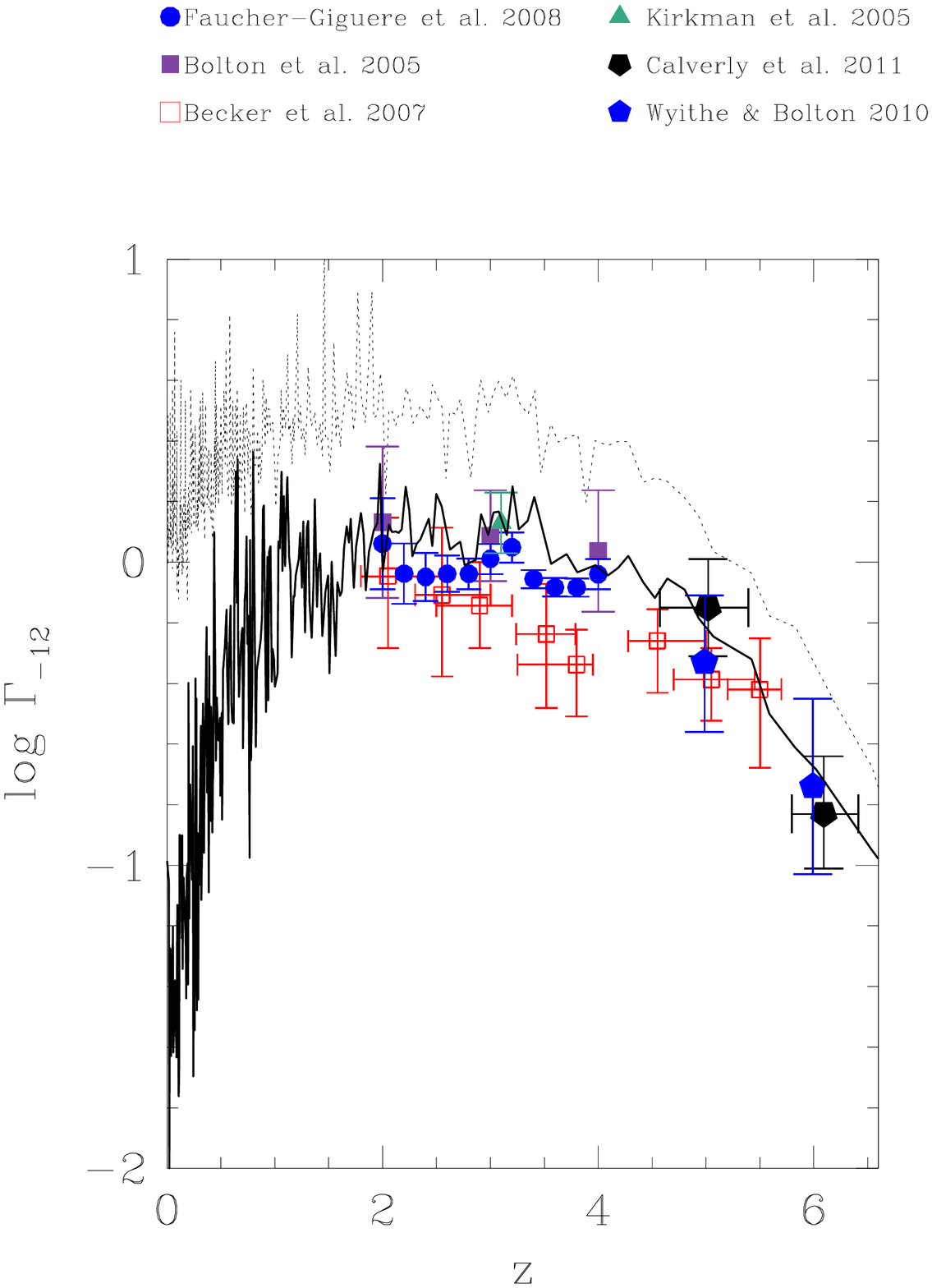}}}
\end{center} {\footnotesize \vspace{-0.2cm }
Fig. 5. - Photoionization rate per hydrogen atom produced by the AGN population in the R-SAM model as a function of redshift is compared with data from different authors scaled to the same IGM temperature-density relation as in \citet{calverley11}. The solid line is the model prediction after eq. 6 and 7, while the dotted one represents the intrinsic production of UV ionizing photons by the overall AGN population (computed from eq. 6 and 7, but assuming an escape fraction $\langle f\rangle=1$ in eq. 6). References for the different data points are 
shown on  top of the figure.
 \vspace{0.4cm}}

\clearpage

\begin{center}
\vspace{-0.5cm}
\scalebox{0.5}[0.5]{\rotatebox{0}{\includegraphics{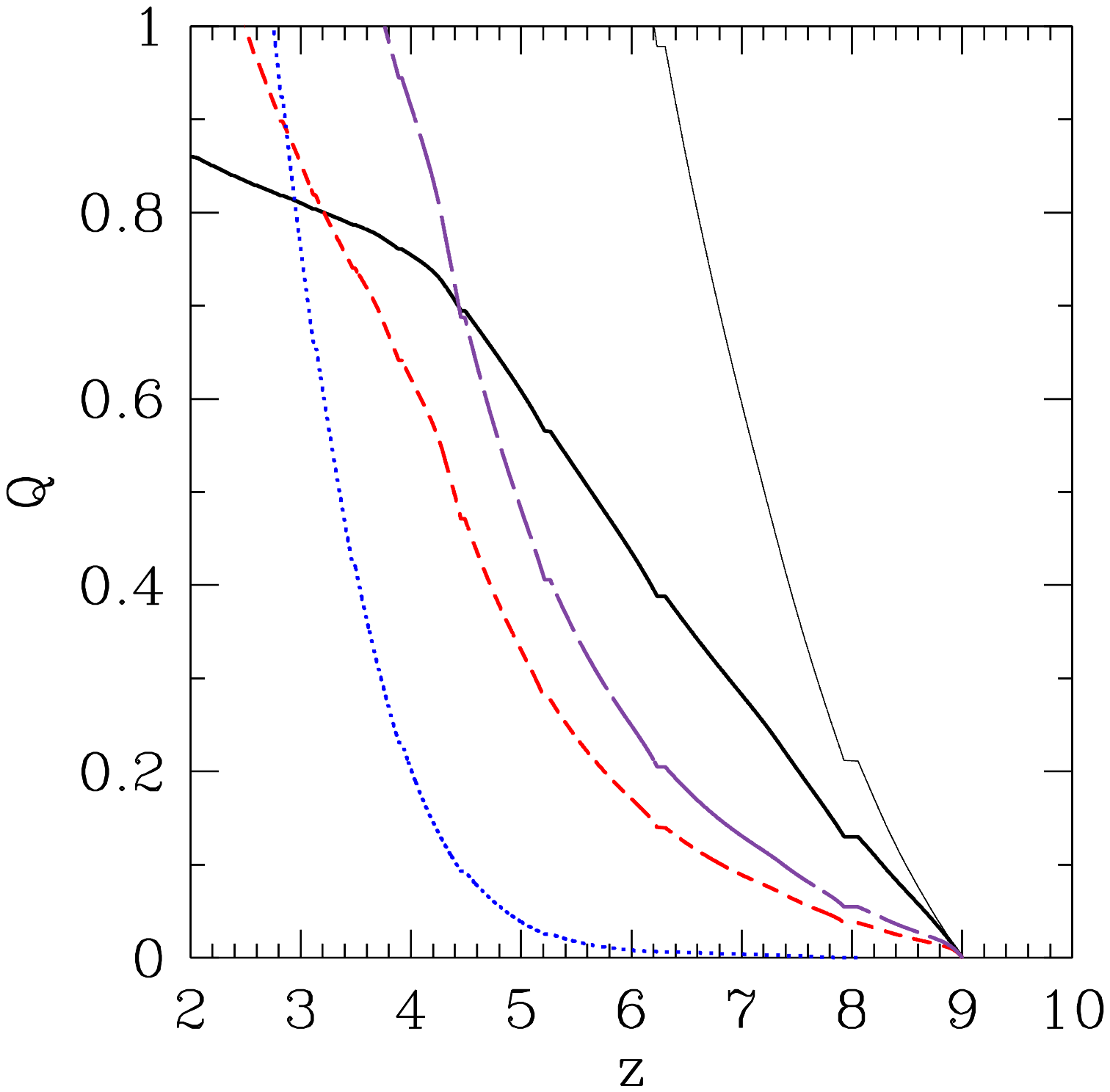}}}
\end{center} {\footnotesize \vspace{-0.2cm }
Fig. 6. - Volume filling factor $Q$ of HeIII and HII as a function of redshift. Dashed (red) curve shows Q(HeIII) with He recombination and clumping factor $C=10$; long dashed (violet) curve shows Q(HeIII) with $C=7$; continuous (black) thick curve also includes photon sink by LLSs and $C=3$; \citet{hama12} is shown as dotted (blue) curve for comparison; thin continuous (black) curve is the HII volume filling factor evolution assuming $C=10$.

 \vspace{0.4cm}}

\end{document}